\documentclass[aps,prx,twocolumn,superscriptaddress,nofootinbib,notitlepage,shortbibliography]{revtex4-1}

\usepackage{amsbsy}
\usepackage{amsfonts}
\usepackage{amsmath}
\usepackage{amstext}
\usepackage{amsthm}
\usepackage{amssymb}
\usepackage{bm}
\usepackage[normalem]{ulem}
\usepackage{color}
\usepackage{comment}
\usepackage{booktabs}
\usepackage{datetime}
\usepackage{dsfont}
\usepackage{dcolumn}
\usepackage{esint}
\usepackage{extarrows}
\usepackage{graphicx}
\usepackage[colorlinks,citecolor=blue]{hyperref} \usepackage{cleveref}
\usepackage{longtable,booktabs}
\usepackage{mathrsfs}
\usepackage{multirow}
\usepackage[super]{nth}
\usepackage{subfigure}
\usepackage{tensor}
\usepackage{url}
\usepackage{xfrac}
\usepackage{xcolor}

\hypersetup{colorlinks=true, linkcolor=blue, filecolor=magenta, urlcolor=blue,}

\begin{document}
\graphicspath{{figures/}}

%\preprint{APS/123-QED}

%\title{Distinct quasiparticle excitations of single-particle spectral function
%Composite quasiparticle structure in local single-electron spectral function

\title{ Composite Structure of Single-Particle Spectral Function in Lightly-Doped Mott Insulators}
%Composite quasiparticle structure in local single-electron spectral function


\author{Jing-Yu Zhao} \affiliation{Institute for Advanced Study, Tsinghua University, Beijing 100084, China} 
\author{Zheng-Yu Weng} \affiliation{Institute for Advanced Study, Tsinghua University, Beijing 100084, China}


%\author{Charlie Author} \homepage{http://www.Second.institution.edu/~Charlie.Author}
%\affiliation{ Second institution and/or address This line break forced with }%
%\affiliation{ Third institution, the second for Charlie Author }%


\date{\today}% It is always \today, today,

\begin{abstract} 
The internal structure of doped holes in the Mott insulator may provide important insight into the physics of doped cuprates. Its observability via a single-particle probe by scanning tunneling spectroscopy (STS) and angle-resolved photo-emission spectroscopy (ARPES) is explored in this paper. Specifically we study the single-particle spectral function based on a two-hole variational ground state wavefunction [Phys. Rev. X 12, 011062 (2022)] in the $t$-$J$ model. The latter as a strongly correlated state possesses a dichotomy of $d$-wave Cooper pairing and $s$-wave ``twisted'' hole pairing. This pairing structure will give rise to two branches of local spectral function at finite energies. The low-lying one corresponds to a nodal-like quasiparticle excitation and the higher branch is associated with the pair breaking of ``twisted'' quasiparticles, with the threshold energy resembling a pseudogap, which is consistent with the recent STS observation.  It can be further extended into energy spectra in momentum space measurable by ARPES, where the low-energy dispersion is also shown to agree well with the Quantum Monte Carlo numerical result for a single hole. It implies that the dominant pairing force arises from the ``twisted'' holes showing up in the high-energy branch. The effect of the next nearest neighbor hopping integral $t'$ is also examined, which shows interesting distinction between $t'/t>0$ and $t'/t\leq 0$ with a dramatic shift of the low-lying excitation from the nodal region to the antinodal region, but with the high-energy branch remaining insensitive to $t'$. Finally, a possible ``orthogonality catastrophe'' effect, namely, a ``dark matter'' component in the strongly correlated wavefunction that cannot be directly detected by the single-electron spectroscopy, is briefly discussed.

%Recent scanning tunneling microscopy (STM) measurements have observed a multi-peak energy structure on the positive bias side ibn dilute-hole-doped cuprates, where tightly-bound hole pairs are identified as the building blocks that can continuously persist into the superconducting regime. 
%In this work, we study the single-particle spectral function based on a two-hole ground state wavefunction [Phys. Rev. X 12, 011062 (2022)], 
%by utilizing variational Monte Carlo method, 
%which can provide a consistent understanding of the experimental results. Here the wavefunction structure with a dichotomy of $d$-wave Cooper pairing and $s$-wave ``twisted'' hole pairing will lead to two branches in the local spectral function where the low-lying one corresponds to a conventional nodal quasiparticle, and a higher energy branch is associated with a ``twisted'' quasiparticle above the pair breaking or ``pseudogap'' energy. These energies can be further extended into energy spectra in the momentum space, in which the low-energy dispersion agrees excellently with the Quantum Monte Carlo numerical result. The implications for the STM spectra in the superconducting state will also be discussed. 

\end{abstract}

%\keywords{Suggested keywords}%Use showkeys class option if keyword display desired
\maketitle

\tableofcontents

\section{Introduction}\label{sec:intro}

The Cooper pair is an essential building block in the Bardeen-Cooper-Schrieffer (BCS) theory of superconductivity. In the cuprate superconductor, the pairing of doped charges is widely believed to exist even above the superconducting (SC) transition temperature $T_c$ in the so-called pseudogap regime, which is short of phase coherence \cite{Emery1995, Lee2006}. Recently, the scanning tunneling spectroscopy (STS) experiments \cite{Cai2016,Ye2023,Li2022,Ye2023b}
have refocused on this issue by detecting local hole pairs in the localized domains/puddles of lightly doped cuprate compounds even in the insulating antiferromagnetic (AF) regime. In particular, the typical STS signature there can continuously persist and evolve into the one with phase coherence in the SC regime. The STS measurements \cite{Ye2023,Li2022,Ye2023b} have further revealed a multiple peak structure on the positive bias side, where the low-lying peak becomes sharpened in the SC phase coinciding with the $d$-wave nodal quasiparticle, while the higher energy peak moves down and evolves to the pseudogap energy scale. Remarkably, in the hole rich region each hole pair seems to occupy a $\sim 4a_0\times 4a_0$ area ($a_0$ is the lattice constant) which remains approximately a stable building block with the sample continuously doped into the SC state at least in the lightly-doping regime.

Experimentally the crucial issue is how a ground state structure can be properly read off with injecting the electron or hole into the system via STS and angle-resolved photo-emission spectroscopy (ARPES). For example, on the positive bias, an electron injected into the local two hole pair region can result in a single hole state, whose detailed energy structure should be closely reflected in the local (STS) spectral function. In a strongly correlated system like the cuprate, how a peculiar pairing structure in the ground state can be explicitly revealed by the STS and ARPES experiments is an important but nontrivial issue even in the two-hole-doped limit because the doped holes here are already strongly entangled with the spin background. 

Theoretically, a recent variational ground-state wavefunction for two holes has shown \cite{Zhao2022} that a tightly-bound pair of doped holes does exist in an AF spin background. It can yield an excellent agreement with the exact diagonalization (ED) and density matrix renormalization group (DMRG) numerical calculations, including the ground-state energy and nontrivial quantum number of angular momentum $L_z=2$ \cite{Zhao2022}. The pairing mechanism is found to have nothing directly to do with the long-range AF fluctuations. Instead, the tightly-bound hole pair is shown to be the natural consequence of removing the surrounding spin currents generated by the hopping of individual single holes by pairing. A peculiar pair structure present in the wavefunction is that it is an $s$-wave pairing in terms of the so-called ``twisted'' hole. The latter is described by a bare hole surrounded by a spin current vortex due to the phase string effect \cite{Wang2015,Chen2019,Zhao2023}. In other words, the ground state is no longer BCS-like as the Landau quasipartices are replaced by the ``twisted holes'' upon doping. On the other hand, a finite overlap of the two-hole ground state with a Cooper pair created on the AF background indicates that the $d$-wave symmetry is naturally realized in the Cooper pair channel of such a two-hole ground state \cite{Zhao2022}. 

Given the pairing symmetry dichotomy in the two-hole ground state, a local spectral function can be calculated to provide valuable information on the internal structure of the wavefunction, which may be directly detected in an STS measurement. Moreover, since two holes are tightly bound in each pair, a condensate of a finite density of the pairs naturally leads to a $d$-wave SC state based on the two-hole wavefunction \cite{Zhao2022}, such that the calculated STS local spectral function may also be used to account for the properties of the SC state, to a leading order of approximation, by assuming that each hole pair as a building block remains robust up to a finite doping as indicated in the experiments \cite{Li2022,Ye2023,Ye2023b,Cai2016}. Furthermore, treating the tightly-bound hole pairs uniformly and coherently distributing in space, one can similarly calculate the spectral function in the momentum space such that the single-hole energy dispersion may be also inferred.      

In this paper, building on such a two-hole ground state ansatz \cite{Zhao2022}, the single-electron spectral function is studied, which shows some very unconventional properties. Specifically a two-component energy structure is revealed, which explicitly corresponds to the pairing structure hidden in the ground state. The low-energy branch shows a conventional quasiparticle excitation (resembling a $d$-wave nodal Bogoliubov quasiparticle) and the higher-energy branch corresponds to a fractionalization of such a quasiparticle into a ``twisted'' hole loosely bound to a spin antivortex field in the background. The onset of the high energy component is found to precisely coincide with the binding energy of two ``twisted'' holes in the ground state, which resembles a precursor of the pseudogap behavior. The momentum distribution functions of the doped holes for the two branches are also examined. As comparison, we have also explored the spectral function on negative bias side by knocking out an electron to create a hole. The excitation dispersion matches with the quantum Monte Carlo result for a single hole remarkably well in the whole Brillouin zone. In particular, it is shown that such a ``free'' single-hole dispersion is symmetric to the low-branch of the positive bias side, except for the vanishing spectral weight in the latter along the nodal lines. Furthermore, the high-energy branch disappears due to an ``orthogonality catastrophe'' effect brought in by the aformentioned antivortex field when the overlap with regard to the half-filling spin background is calculated.        

Moreover, the effect of a next nearest neighbor (NNN) hopping integral $t'$ on the spectral function is also examined. 
At $t'/t\leq 0$, the low-lying branch of nodal quasiparticle remains at the nodal regions centered around the momenta $(\pm \pi/2,\pm\pi/2)$ with merely a change in the anisotropic equal-energy contours. But an interesting shift of the quasiparticle spectrum from the nodal to the antinodal region occurs at $t'/t>0$ beyond a small critical value. By contrast, the higher energy branch remains robust to the perturbation of $t'$ with negligible change. In spite of such a surface change in the low-lying quasiparticle spectroscopy, the underlying two-hole ground state structure or pairing mechanism has no essential change as a function of $t'$.

Finally, in the summary and discussion section, we briefly discuss the so-called ``orthogonality catastrophe'' in the single-electron probes by STS and ARPES. We point out that an important portion of the strongly correlated wavefunctions, involving the spin currents generated by doping in the spin background, may not be observed directly in these experiments with vanishing spectral weight. An example of a localized hole by impurity is discussed, in which the STS can fail to detect a local uniform distribution of the hole, with an observation of a clover-like pattern of finite spectral weight instead.  It may thus complicate the theoretical understanding of the experimental measurements made in the cuprate, which is widely believed to be a doped Mott insulator \cite{Anderson1987,Lee2006}.

\section{Model and variational wavefunctions}
The $t$-$J$ model is a minimal model describing the doped Mott insulator physics in the cuprate \cite{Anderson1987,Lee2006}: 
\begin{equation} \label{eqn:tjH}
    H =-\sum_{\langle ij\rangle,\sigma}t(c_{i\sigma}^\dagger c^{}_{j\sigma}+h.c.)+
    \sum_{\langle ij\rangle}J\left(\mathbf{S}_i\cdot\mathbf{S}_j-\frac{1}{4}n_in_j\right), 
\end{equation}
where $\langle ij\rangle$ denotes a nearest-neighbor (NN) bond, and the strong correlation nature originates from the no double occupancy constraint $\sum_{\sigma} c^\dagger_{i\sigma}c^{}_{i\sigma}\leq 1$. Here 
$\mathbf{S}_i$ and $n_i$ are spin and electron number operators on site $i$, respectively. We choose $t/J=3$ throughout the paper. 

%\section{Variational wavefunction ansatz}\label{sec:model}
%\subsection{Twisted hole as elementary quasi-particle.}

%\textit{Variational wavefunctions.---}
The $t$-$J$ model reduces to the Heisenberg spin model at half-filling, where the ground state $|\phi_0\rangle$ is a spin-singlet state for a finite-size sample. Here $|\phi_0\rangle$ can be very precisely simulated by a bosonic resonating-valence-bond (RVB) wavefunction \cite{Liang1988,Chen2019}, which produces a long-range antiferromagnetic (AF) spin correlation in the large sample limit.  
%Moving forward, we inject a hole at site $i$ by removing a spin-$\downarrow$ electron from the spin singlet background. 
%Naively, this will give rise to a set of single-hole basis $c_{i\downarrow}|\phi_0\rangle$. 

%%%%%%%%%%%%%%%%%%%%%%%%%%%%%%%%%%%%%%%%%%
\begin{figure}[t]
    \centering
    \includegraphics[width=0.46\textwidth]{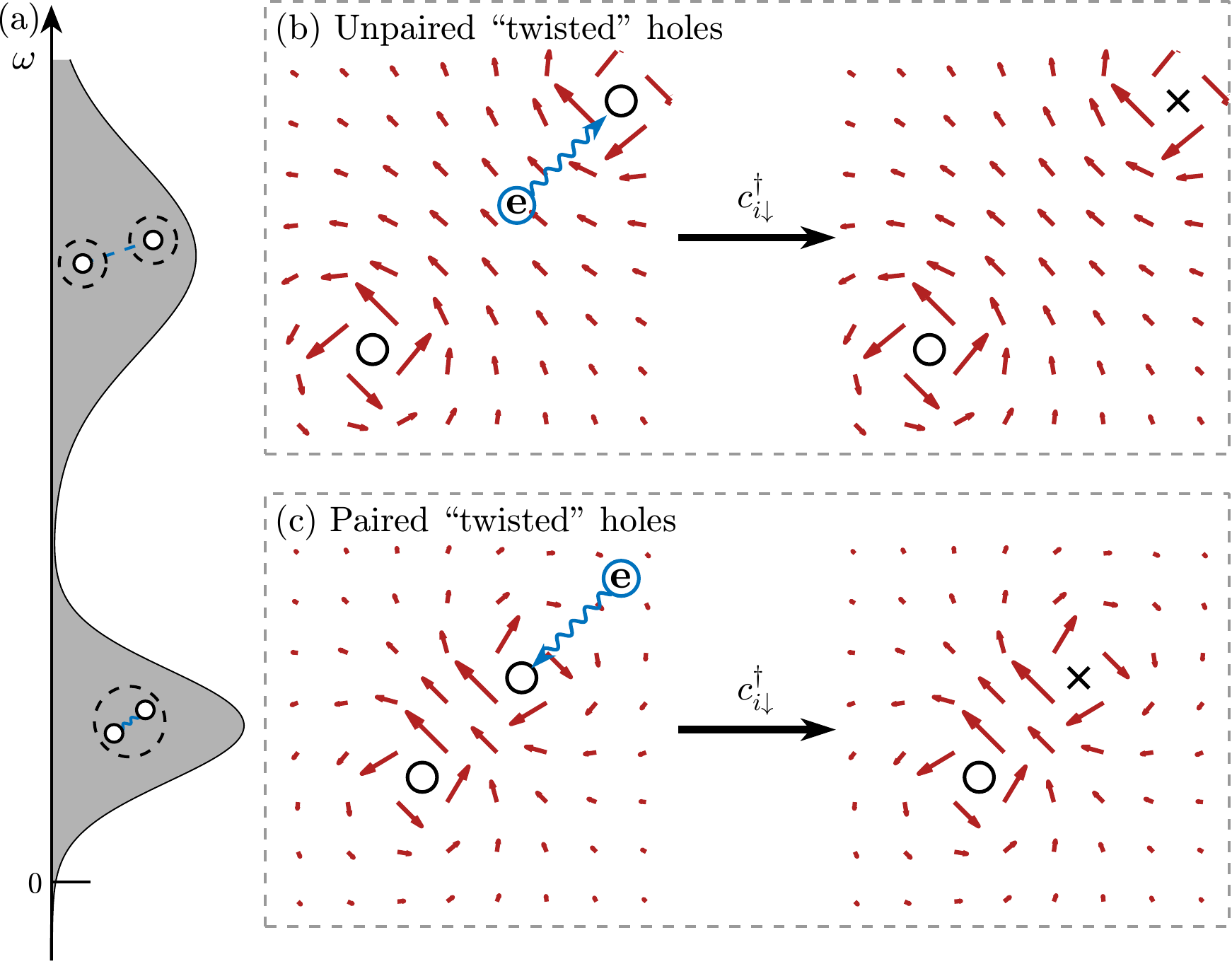}
     \caption{Schematic illustration of the local single-particle spectral function probing into the two-hole ground state, which exhibits a double peak structure in (a): The high-energy peak corresponds to a single-hole excitation via a process by first unbinding the ``twisted'' hole pair and then annihilating one of the holes with an injecting bare electron in (b); The lower-energy peak corresponds to directly annihilating one bare hole of the tightly paired holes in (c). Since the single-hole excitation here is created by injecting a bare electron into the two-hole ground state via $c^{\dagger}_{i\sigma}$, it naturally inherits the two-component structure of the latter \cite{Zhao2022}. The vortex-antivortex patterns of spin currents shown by red arrows are loosely and tightly bound in (b) and (c), respectively, indicating the single-hole excitation is either fractionalized into a ``twisted'' hole separated from an antivortex or confined with the latter to recover the Landau quasiparticle (see text). }

   %  and the ``twisted'' hole is created by $\tilde{c}_{i\sigma} = e^{-i\hat{\Omega}_i}c_{i\sigma}$, where $ e^{-i\hat{\Omega}_i}$ describes a transverse spin current vortex [cf. Eq.~(\ref{eqn:phasestringoperator})] around the bare hole (circle) or the empty center (cross) [cf. the patterns of red arrows in (b) and (c)]. []  }
 %   \caption{(a) Schematic illustration of the two peaks in the local single-particle spectral function, in which the higher-energy peak is at the energy that two holes are first unpaired in the intermediate state and then one of them gets annihilated by injecting electron as indicated by (b), and the lower-energy peak corresponds to directly annihilating one of the two holes in the paired state (c); Here (b) and (c) show the corresponding spin current patterns (red arrows) associated with both processes with the black circle and cross denoting the hole and the core of the anti-vortex left by the annihilated hole, respectively (see text). 
        %For the low-energy peak, the two-hole state is in its confined ground state. 
        %Upon introducing a bare electron into the system, the spin current vortex , leaving a strongly paired hole and anti-vortex. 
        %On the other hand, for the high-energy peak where the holes are deconfined, the corresponding single hole state will be a deconfined twisted-quasiparticle.  
        
    \label{fig:illu}
\end{figure}
%%%%%%%%%%%%%%%%%%%%%%%%%%%%%%%%%%%%%%%%%%%

\subsection{``Twisted'' quasiparticle }\label{T}
%\subsection{Single-hole wavefunction.}

Based on $|\phi_0\rangle$, a single-hole wavefunction has been recently constructed \cite{Wang2015,Chen2019,Zhao2023}. Instead of a bare hole state $c_{i\sigma}|\phi_0\rangle$ with removing a spin-$\sigma$ electron from the half-filling background, the ground state ansatz takes the following form \cite{Chen2019} %\cite{Weng2011a,Wang2015,Chen2019}. 
\begin{equation}\label{eqn:singleholeansatz}
    |\Psi_G\rangle_{1h} = \sum_{i} \varphi(i)\tilde{c}_{i\sigma}|\phi_0\rangle~,
\end{equation}
where a ``twisted'' hole creation operator is given by 
\begin{equation}\label{twisted}
\tilde{c}_{i\sigma} = e^{-i\hat{\Omega}_i}c_{i\sigma}~,
\end{equation}
and $\varphi(i)$ is the wavefunction of such a twisted hole determined by VMC.
Here the specific expression of the phase factor $\hat{\Omega}_i$ is defined by
\begin{equation}\label{eqn:phasestringoperator}
    \hat{\Omega}_i = \sum_{l(\neq i)}\theta_i(l)n_{l\downarrow}~,
\end{equation}
where $n_{l\downarrow}$ denotes the spin-$\downarrow$ number operator of the electron at site $l$, and $\theta_i(l)$ is the so-called statistical angle $\mod 2\pi$ defined by the angle between the arrow pointing from site $l$ to $i$ and the horizontal line.

Once employing the ``twisted'' electron $\tilde{c}_{i\sigma}$ in Eq.~\eqref{eqn:singleholeansatz}, the
VMC calculations can yield good agreements with ED and DMRG results on both finite-size 2D square lattice  \cite{Zheng2018b,Chen2019,Chen2022} and two-leg ladder \cite{Wang2015,Zhao2023,Zhu2015,Zhu2015b}, 
including an exotic quantum number, i.e., the orbital angular momentum $L_z=\pm 1$ in the 2D ground state. 
Here the ``twisted'' quasiparticle $\tilde{c}_{i\sigma}$ can be identified with a composite structure comprised of a bare hole with a chiral spin current pattern surrounding around it \cite{Chen2019,Chen2022}. Equation ~\eqref{eqn:singleholeansatz} may be rewritten as 
\begin{equation}\label{eqn:unitary}
    |\Psi\rangle_{1h} = e^{i\hat{\Theta}} \sum_{i} \varphi(i)c_{i\sigma}|\phi_0\rangle~,
\end{equation}
which differs from a conventional Laudau's quasiparticle state $\sum_{i} \varphi(i)c_{i\sigma}|\phi_0\rangle~$ by a duality transformation 
\begin{equation}
e^{i\hat{\Theta}}\equiv e^{-i\sum_l n_l^h\hat{\Omega}_l}
\end{equation}
with $n_l^h$ denoting the hole number operator at site $l$. In a large sample size limit, one has \cite{Zhao2022}
\begin{equation}\label{Omega}
\langle\phi_0|e^{i\hat{\Omega}_j}e^{-i\hat{\Omega}_i}|\phi_0\rangle
\end{equation}
vanishing exponentially with $|i-j|\gg 1$ in 2D \cite{Zhao2022}, such that there is an ``orthogonality catastrophe'' for the single-hole state Eq.~\eqref{eqn:singleholeansatz} with the quasiparticle spectral weight $Z_k\rightarrow 0$ \cite{Chen2019}. On the other hand, for the two-leg ladder with a spin gap in the background $|\phi_0\rangle$, $e^{-i\hat{\Omega}_i}$ exhibits a quasi-long-range order such that $Z_k \neq 0$ \cite{Zhao2023}. Nonetheless, the ``twisted'' single-hole state Eq.~\eqref{eqn:singleholeansatz} is shown \cite{Zhao2023} to be a non-Landau quasipartilce in this case with vanishing charge to effectively become a neutral spinon.      

In general, the feedback of the ``twisting'' hole on the spin background has to be incorporated via 
\begin{equation}\label{RVB}
|\phi_0\rangle\rightarrow |\mathrm{RVB}\rangle
\end{equation}
which is neglected in Eqs.~\eqref{eqn:singleholeansatz} and (\ref{eqn:unitary}) as a mobile single-hole effect is of order of O(1/N) for a finite-size $N$ system. According to $c_{i\sigma}=\tilde{c}_{i\sigma}e^{i\hat{\Omega}_i}$, self-consistently the bare hole state should be renormalized into $c_{i\sigma}|\phi_0\rangle\rightarrow \tilde{c}_{i\sigma}|\mathrm{RVB}\rangle$ with the antivortex $e^{i\hat{\Omega}_v}$ incorporated into $|\mathrm{RVB}\rangle$ near the hole, whose form is to be explicitly given later following the introduction of the two-hole ground state below. 
%Note that a singular effect arising due to the long-range AFM order in the thermodynamic limit of 2D will need a special attention as to be elaborated elsewhere. Here we always focus on a finite-size system to explore the hopping process which is optimized via Eq. (\ref{twisted}). On the other hand, to recover a Landau quasiparticle excitation, created by $c_{i\sigma}=\tilde{c}_{i\sigma}e^{i\hat{\Omega}_i}$ in the spectral function calculated later, one has to incorporate a local anti-twist $e^{i\hat{\Omega}_v}$ in $|\mathrm{RVB}\rangle$ near the hole, which is to be further discussed following the introduction of the two-hole ground state below.  

\subsection{Two-hole ground state wavefunction}

Based on $|\phi_0\rangle$, a two-hole variational ground state  has been recently constructed,
which is described by the wavefunction ansatz \cite{Zhao2022}:
\begin{equation} \label{eqn:twoholeansatz}
    |\Psi_G\rangle_{2h} = \sum_{i,j} g(i,j)e^{-i(\hat{\Omega}_i-\hat{\Omega}_j)}c_{i\uparrow}c_{j\downarrow}|\phi_0\rangle~,
\end{equation}
with $g(i,j)$ as the variational parameter. The wavefunction in Eq.~\eqref{eqn:twoholeansatz} has been extensively studied by the VMC method in Ref.~\onlinecite{Zhao2022} with an excellent agreement with the exact numerics including the new quantum number $L_z=2$ in the ground state. A dichotomy of the pairing symmetry has been identified here \cite{Zhao2022}: while $g(i,j)$ is $s$-wave-like with a strong NN pairing amplitude, $|\Psi_G\rangle_{2h}$ is shown to be $d$-wave-like as measured by the Cooper pair order parameter $c_{{\bf k}\uparrow}c_{-{\bf k}\downarrow}$. The strong pairing as evidenced by a big unpairing energy scale $E_{\mathrm{pair}}\sim 1.97J$ has been found \cite{Zhao2022} due to the cancellation between the spin currents of opposite chiralities in Eq.~\eqref{eqn:twoholeansatz}. 
Namely the ``twisted'' hole object in Eq.~\eqref{eqn:singleholeansatz} is strongly frustrated, whose kinetic energy can only get restored by pairing in Eq.~\eqref{eqn:twoholeansatz}, which has been pointed out to be a new pairing mechanism fundamentally distinct from the usual RVB pairing mechanism.    

It has been argued \cite{Zhao2022} that the tightly-bound two holes in Eq.~\eqref{eqn:twoholeansatz} form a ``building block'' of size $\sim 4a_0\times 4a_0$, which eventually leads to a superconducting domain or phase at finite-doping. In other words, the Bose condensate of the two-hole state Eq.~\eqref{eqn:twoholeansatz} may be a good candidate for the superconducting ground state \cite{Weng2011a} when a finite density of holes are present. To be generalized to finite doping, Eq.~\eqref{eqn:twoholeansatz} may be also reexpressed as 
\begin{equation} \label{eqn:twohole}
    |\Psi_G\rangle_{2h} = e^{i\hat{\Theta}} \sum_{i,j} g(i,j)c_{i\uparrow}c_{j\downarrow}|\mathrm{RVB}\rangle~,
\end{equation}
where the variational half-filling spin background takes a form \cite{Zhao2022} $|\mathrm{RVB}\rangle \sim \left(e^{i2\hat{\Omega}_i}+e^{i2\hat{\Omega}_j}\right)|\phi_0\rangle$.
At finite doping of $N_h$ holes, the SC ground state has been proposed \cite{Weng2011a} as the condensate of the hole pairs described in Eq.~(\ref{eqn:twohole}) as follows
\begin{align} \label{eqn:SC}
    |\Psi_G\rangle = e^{i\hat{\Theta}} \left\{\sum_{i,j} g(i,j)c_{i\uparrow}c_{j\downarrow}\right\}^{N_h/2}|\mathrm{RVB}\rangle \nonumber \\
     \propto  \left(\hat{\cal D}\cdot\hat{\cal D}\cdot... \hat{\cal D}\right) |\mathrm{RVB}\rangle~, \ \ \ \ \ \ \ \ \ \
\end{align}
with $\hat{\cal D}\equiv \sum_{i,j} g(i,j)\tilde{c}_{i\uparrow}\tilde{c}_{j\downarrow}$ and $|\mathrm{RVB}\rangle$ evolving from $|\phi_0\rangle$ self-consistently \cite{Weng2011a,Ma2014}.
The pairing symmetry dichotomy is shown \cite{Ma2014} by that while $g(i,j)$ remains $s$-wave-like, the Cooper pairing order parameter is $d$-wave-like, similar to the two-hole state in Eq.~\eqref{eqn:twoholeansatz} or (\ref{eqn:twohole}).  

%Since a local two-hole ground state in Eq.~\eqref{eqn:twoholeansatz} may thus be well probed by the STM experiment, which amounts to the calculation of the local single-particle spectral function to be studied below.      

%By contrast, if $\tilde{c}_{i\sigma}$ in Eq.~\eqref{eqn:singleholeansatz} is replaced by a bare electron operator $c_{i\sigma}$, none of the exotic results above can be obtained in the VMC procedure, indicating the ``twisted'' hole state in Eq.~\eqref{twisted} that the correct low-lying excitation differs from the bare  as the coherent single particle excitation. 

\subsection{The single-hole excitation state and the spectral function} 

In the above, we have outlined the VMC study  in which the two-hole ground state ansatz in Eq.~(\ref{eqn:twoholeansatz}) has been established as the building block properly describing the doping effect in the $t$-$J$ model \cite{Zhao2022}. In the following, we examine how such a local pairing of holes can be probed by the single-particle measurements like STS and ARPES. It may be generalized to finite doping as to be discussed later.  

To probe the local two-hole ground state by, say, STS measurement, a single-hole excitation can be constructed based on, say, $c^{\dagger}_{i\uparrow}|\Psi_G\rangle_{2h}$ according to Eq. (\ref{eqn:twoholeansatz}), which may be generally expressed by 
\begin{equation}\label{eqn:singleholeansatz_v}
    |\Psi^c\rangle_{1h} = \sum_{i,v} \varphi(i,v)e^{-i(\hat{\Omega}_i-\hat{\Omega}_v)}c_{i\downarrow}|\phi_0\rangle~,
\end{equation}
where, without loss of generality, a hole with $\downarrow$-spin is considered. Here $|\Psi^c\rangle_{1h}$ replaces Eq.~\eqref{eqn:singleholeansatz} by an antivortex $e^{i\hat{\Omega}_v}$ created on $|\phi_0\rangle$ with an amplitude $\varphi(i,v)$ to give rise to Eq.~(\ref{RVB}), with  $\varphi(i,v)$ determined by the VMC method \cite{Zhao2022}, where the center $v$ relaxed from a site to a plaquette.

Thus, an electron (with spin-$\uparrow$) injected into a local two-hole paired state, $|\Psi_G\rangle_{2h}$, in an STS probe, can be described by a local single-particle spectral function at site $i$ on the positive bias side ($\omega\geq 0$) by
\begin{equation} \label{eqn:Ap}
    A_{ii}^p(\omega) = -\mathrm{Im} \sum_n\frac{\left|_{1h}\langle\Psi^c(n)|c^\dagger_{i\uparrow}|\Psi_G\rangle_{2h}\right|^2}
    {\omega-[E_{1h}(n)-E^G_{2h}-\mu_p] +i\eta}~,
\end{equation}
where $|\Psi^c(n)\rangle_{1h}$ denotes the single-hole excitation state in Eq.~\eqref{eqn:singleholeansatz_v}
with energy spectrum $E_{1h}(n)$. Here $E^G_{2h}$
denotes the two-hole ground state energy and the chemical potential $\mu_p$ is set to make the lowest energy of $E_{1h}(n)$ shift to $\omega=0$ such that $A_{ii}^p(\omega)\neq 0$ at $\omega\geq 0$ at the broadening $\eta\rightarrow 0^+$. 
Note that in the two-hole ground state Eq.~\eqref{eqn:twoholeansatz}, each of the two holes carries a meron-vortex with opposite chiralities. Once an electron is injected into the system to annihilate the hole, the associated meron-vortex is left by a sudden approximation, which accounts for the difference between a ``twisted'' hole and a bare hole according to Eq. (\ref{twisted}). 

On the negative bias side, the STS can further probe the single-hole excitation $|\Psi^c(n)\rangle_{1h}$ via, e.g.,  $c_{i\downarrow}|\phi_0\rangle$, as follows 
\begin{equation} \label{eqn:Am}
    A_{ii}^n(\omega) = -\mathrm{Im} \sum_n\frac{\left|_{1h}\langle\Psi^c(n)|c_{i\downarrow}|\phi_0\rangle\right|^2}
    {\omega +[E_{1h}(n)-E^G_{0}-\mu_n]+i\eta}~,
\end{equation}
where $E^G_{0}$ denotes the ground state energy of  
$|\phi_0\rangle$  and the chemical potential $\mu_n$ is chosen such as to move the excitation edge to zero (i.e., $\omega\leq 0$).  In principle, at the same location probed by $A_{ii}^p(\omega)$ at $\omega\geq 0$,  an STS probe at $\omega\leq 0$ should involve
an overlap between a three-hole excited state and the two-hole ground state by knocking one additional electron out of the system, which would be beyond the capability of the current method.  By assuming the newly created hole is only weakly coupled to the two-hole pair in the background, we may compute $A_{ii}^n(\omega)$ defined in Eq.~\eqref{eqn:Am} instead, which actually describes the STS probe into a local region in the absence of the doped holes.  

\begin{figure}[t]
    \centering
    \includegraphics[width=0.45\textwidth]{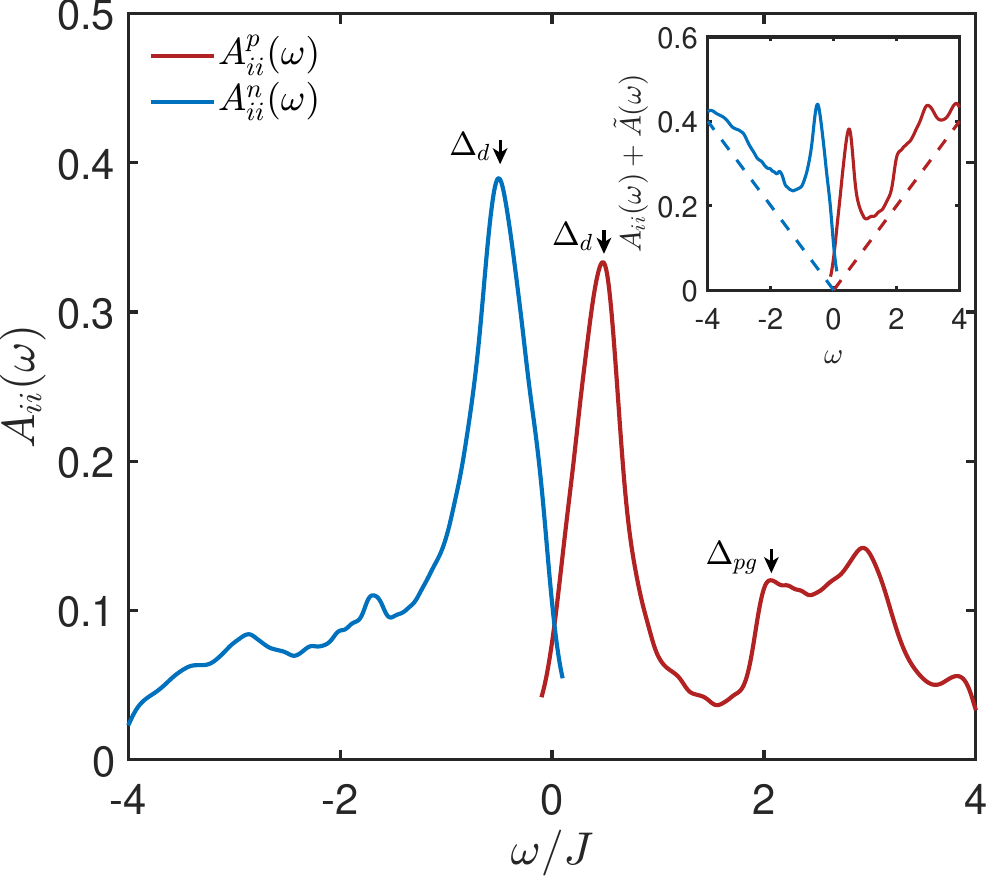}
    \caption{Local single-electron spectral function probed by STS experiment. The red curve on the positive bias ($\omega\geq 0$) side is calculated by injecting an electron into the two-hole ground state as given in Eq.~\eqref{eqn:Ap}; The blue curve at $\omega\leq 0$ is determined by extracting an electron from the half-filling ground state as given in Eq.~\eqref{eqn:Am}. The inset shows the same data with a $V$-shaped background (dashed lines) added for comparison with the experiment data (see text). 
    %The data is averaged over the lattice to remove inhomogeneous. 
    }
    \label{fig:stm_ii}
\end{figure}

Here $A^p_{ii}(\omega)$ and $A_{ii}^n(\omega)$ both involve the single-hole excitation $|\Psi^c(n)\rangle_{1h}$ as given in Eq.~\eqref{eqn:singleholeansatz_v}.   In the following, we can make a qualitative discussion on some of the key features to be expected from $A^p_{ii}(\omega)$, which are related to the novel pairing structure in the ground state Eq.~\eqref{eqn:twoholeansatz}. Figure~\ref{fig:illu} (a) illustrates a double-peak structure of $A^p_{ii}(\omega)$ to be calculated in the next section (cf. Fig.~\ref{fig:stm_ii}). It corresponds to two distinct branches of the energy spectrum of the single-hole excitation $|\Psi^c(n)\rangle_{1h}$. According to the two-hole ground state, the high- and low-energy branches involve the processes illustrated in  Figs.~\ref{fig:illu} (b) and (c),  respectively. In Fig.~\ref{fig:illu} (b), two ``twisted'' holes, which form a tightly bound state in the ground state, become unbound first and then with one hole annihilated by the injected electron. In Fig.~\ref{fig:illu} (c), a ``twisted'' hole remains tightly bound with the antivortex left by the other hole being annihilated. The latter is close to a conventional quasiparicle with the confinement of the ``twisted'' hole and antivortex, whereas the former behaves like a deconfined ``twisted'' hole given in Eq.~(\ref{twisted}) with a loosely bound antivortex. In the next section, we shall use the VMC method \cite{Zhao2022} to determine the excitation energy spectrum of the wavefunction in Eq.~\eqref{eqn:singleholeansatz_v} and examine the spectral functions in detail.

%%%%%%%%%%%%%%%%%%%%%%%%%%%%%%%%%%%%%%%%%%
%%%%%%%%%%%%%%%%%%%%%%%%%%%%%%

\begin{figure}[t]
    \centering
    \includegraphics[width=0.46\textwidth]{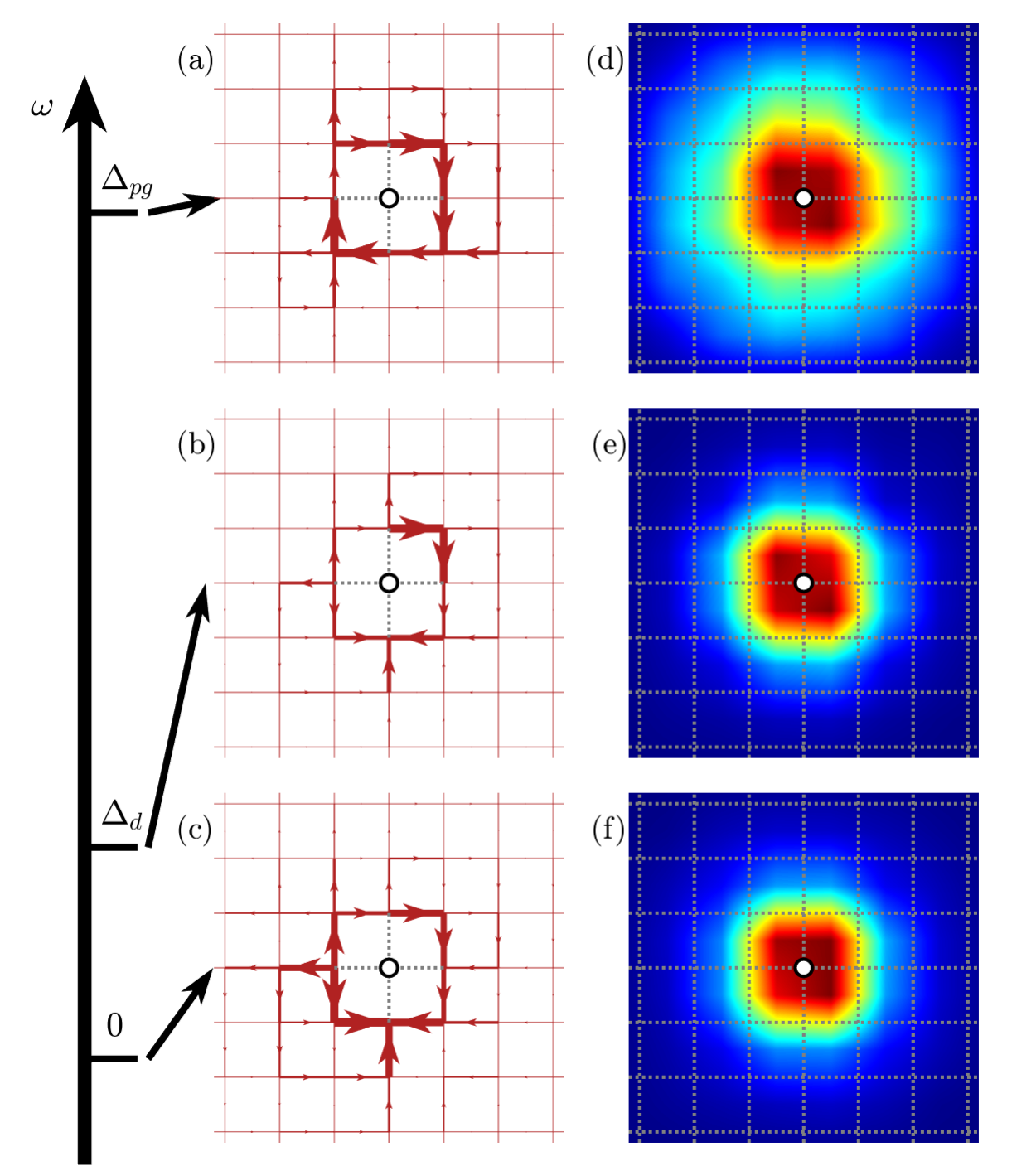}
    \caption{The real-space structures of the single hole excitation in the calculated spectral function at different energy scales as shown in Fig.~\ref{fig:stm_ii} at $\omega>0$. 
        Here (a)-(c) show the local spin current patterns with the hole projected at a given site $h_0$. The thickness of the red arrow represents the strength of the transverse spin current as illustrated in Fig.~\ref{fig:illu};
        (d)-(f) show the antivortex distribution $|\varphi(h_0,v)|$ around the hole.
         }
        %The specific energy scales are chosen as $\Delta_d \sim 0.5J$ and $\Delta_{pg}\sim 2.0J$ in practical calculation. }
    \label{fig:local_spincur}
\end{figure}
%\textit{Single-electron spectral functions.---}
\section{Single-particle spectral functions }

%%%%%%%%%%%%%%%%%%%%%%%%%%%%%%%%%%%%%%%%%%%%%%%
\begin{figure*}[t]
    \centering
    \includegraphics[width=1.0\textwidth]{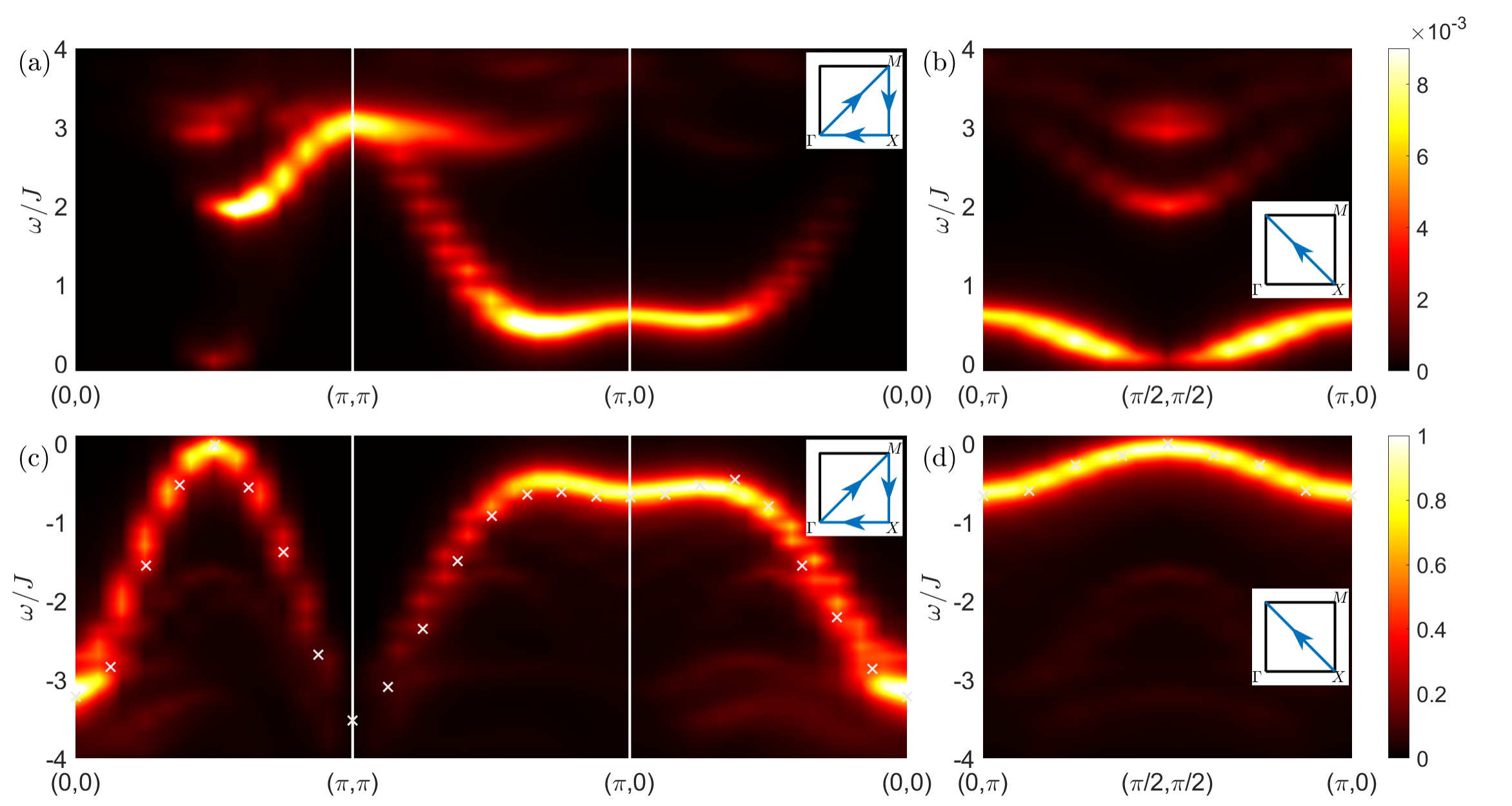}
    \caption{Momentum space single-electron spectral function as calculated by VMC. 
        Here (a) and (b) correspond to injecting an electron of momentum ${\bf k}$ into the two-hole ground state as given in Eq.~\eqref{eqn:Akp};
        (c) and (d) are determined by extracting an electron of momentum $-{\bf k}$ from the half filling ground state as given in Eq.~\eqref{eqn:Akm}. 
        The result of the Green's function Monte Carlo calculation \cite{Boninsegni1994} of the single-hole spectrum is also presented in (c) and (d) (white cross) for comparison. Note that  (a) and (b) share the same color bar, as do (c) and (d), and the momentum cut lines in the Brillouin zone are indicated in the corresponding insets. }
        
    \label{fig:arpes}
\end{figure*}
%%%%%%%%%%%%%%%%%%%%%%%%%%%%%%%%%%%%%%%%%%%%%%%
%%%%%%%%%%%%

\subsection{The STS (local) spectral function}

The local single-particle spectral function $A^p_{ii}(\omega)$ given in Eq. ~\eqref{eqn:Ap} can be computed by the VMC method \cite{Zhao2022}, which is shown in Fig.~\ref{fig:stm_ii} by the red curve at $\omega \geq 0$. Indeed a double-peak structure with two characteristic energies, $\Delta_d\sim 0.5J$ and $\Delta_{pg}\sim 2.0J$ (with $\eta\sim 0.1J$) has been found, in agreement with the previous qualitative discussion (cf. Fig.~\ref{fig:illu}). The spatial structures associated with the underlying wavefunctions will be further examined below. It is noted that such two-branch spectral function is also qualitatively consistent with the STS observation involving local hole pair in the lightly-doped cuprate \cite{Li2022}. It is noted that the overall strength of the spectral function is determined as follows: it is first averaged over and scaled to a typical two-hole size $4a_0\times 4a_0$ \cite{Zhao2022} and then multiplied by a numerical factor 3 as experimentally the STS spectral function is measured at the brightest peak within a $4a_0\times 4a_0$ plaquette structure \cite{Li2022}.

For comparison, the local spectral function at $\omega<0$ given by $A_{ii}^n(\omega)$ in Eq. (\ref{eqn:Am}) is shown by the blue curve in Fig.~\ref{fig:stm_ii}. 
Note that the relative strength between the red and blue curves of $A_{ii}^p(\omega)$ and $A_{ii}^n(\omega)$  is arbitrary in Fig.~\ref{fig:stm_ii}. In the inset of Fig.~\ref{fig:stm_ii}, a V-shaped background (marked by the dashed lines) is added to guide the eye as possible contributions from the background spin-wave excitations, etc., during the tunneling process.

%\subsection{Real-space textures}

%In the single-electron probe like STS and ARPES experiments, the spectral function is determined via the excitations created, say, by $c^{\dagger}$ acting on the ground state. In the present work, we mainly examine the single-hole excitation $|\Psi^c(n)\rangle_{1h}$ constructed based on the two-hole ground state $|\Psi_G\rangle_{2h}$ as given in Eq.~\eqref{eqn:singleholeansatz_v}. In the above, two energy branches in $|\Psi^c(n)\rangle_{1h}$ have been found based on the VMC calculation. In the following, 

Now we further explore the local spin texture surrounding the hole in the excitation $|\Psi^c(n)\rangle_{1h}$ created by $c^{\dagger}_{i\sigma}|\Psi_G\rangle_{2h}$ as probed by $A_{ii}^p(\omega)$. Let us first consider the low-energy branch at $\omega\sim 0$ and $\omega \sim \Delta_d$ [cf. Fig.~\ref{fig:stm_ii} and Fig.~\ref{fig:illu} (c)]. By projecting the hole at a given position $h_0$, the spin current patterns are shown in Figs.~\ref{fig:local_spincur} (c) and (b), respectively. Here in the low-energy branch, the spin current vortex associated with the ``twisted'' hole gets essentially compensated by the antivortex $e^{i\hat{\Omega}_v}$ with the amplitude $\varphi(h_0,v)$ around the hole side $h_0$, which are tightly bound as illustrated in Fig.~\ref{fig:illu}(c) as well as in Figs.~\ref{fig:local_spincur} (f) and (e). Thus, the total spin current gets compensated on average such that the single hole state has an overlap with a conventional quasiparticle measured by the spectral function. 

Then at the high-energy branch with $\omega\sim \Delta_{pg}$ (cf. Fig.~\ref{fig:stm_ii}), the neutral spin current pattern around the hole will adopt the shape like a closed vortex profile [cf. Fig.~\ref{fig:local_spincur}(a)] with the antivortex $e^{i\hat{\Omega}_v}$ well separated from the ``twisted'' hole [cf. the distribution $|\varphi(h_0,v)|$ in Fig.~\ref{fig:local_spincur}(d)]. It corresponds to the configuration as if 
the two holes are close to be unpaired before one hole gets annihilated by the injecting electron in $c^{\dagger}_{i\sigma}|\Psi_G\rangle_{2h}$, which results in a high-energy single-hole state with a ``twisted'' hole loosely bound to an anitvortex as illustrated in Fig.~\ref{fig:illu}(b). Indeed it is consistent with the finding that the energy scale $\Delta_{pg}\simeq E_{\mathrm{pair}}$ - the unbinding threshold for the two-hole pairing state \cite{Zhao2022}. 

%Finally, it is worth to point out that the distinction between Eq.~\eqref{eqn:singleholeansatz}  and $|\Psi^c(n)\rangle_{1h}$ in Eq.~\eqref{eqn:singleholeansatz_v} represents an important fact that injecting a bare electron or hole into the two-hole or half-filled ground state, under a ``sudden approximation'' in the STM and ARPES experiments, does not necessarily creates a true single-hole ground state, which marks a fundamental paradigm shift in understanding the single-particle spectral function for a Mott insulator. Since a bare electron (hole) involves the process of $c_{i\downarrow}\Longleftrightarrow\tilde{c}_{i\downarrow} e^{-i\hat{\Omega}_i}$, a vortex field $e^{i\hat{\Omega}_v}$ will be left in the background as a many-body effect, which effectively prevents Eq.~\eqref{eqn:singleholeansatz_v} from becoming the ``twisted'' single-hole state in Eq.~\eqref{eqn:singleholeansatz} unless a very long-time evolution is realized. The related issue of ``orthogonality catastophe'' involving the relaxation of the antivortex field will need a separate investigation elsewhere. 

%By further analysing at different energy scales, one may gain more insights into the underlying physics of the single-hole state. For example, the local spatial structure surrounding the doped hole in Eq.~\eqref{eqn:singleholeansatz_v} is calculated, as shown in Fig.~\ref{fig:local_spincur}, which offers a view of the surrounding neutral spin current as well as 

 %%%%%%%%%%%%%%%%%%%%%%%%%%%%%%%%%%%%%%%%%%%%%%%%%%%%%%%%%%
\begin{figure}
    \centering
    \includegraphics[width=0.48\textwidth]{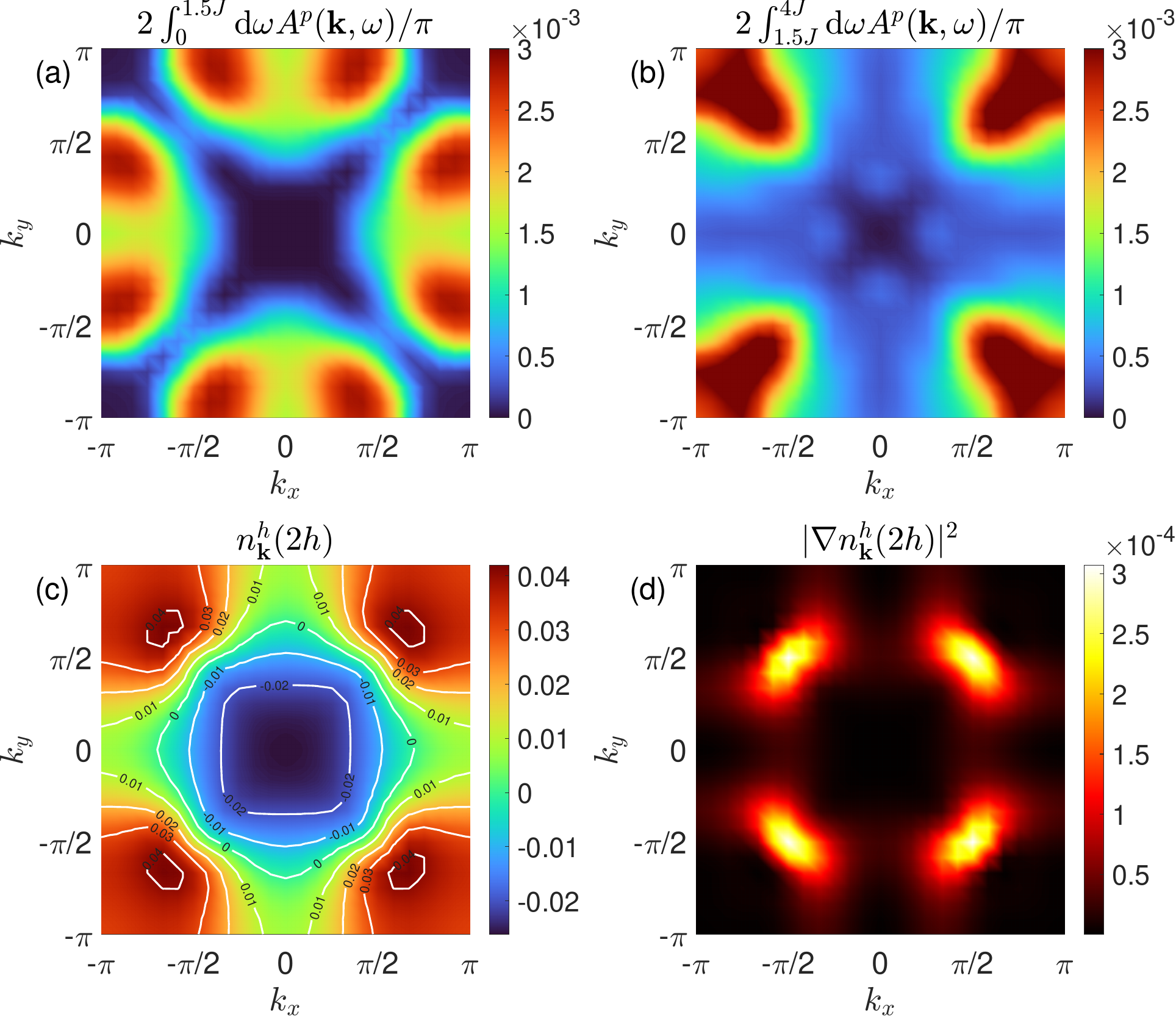}
    \caption{The integrated spectral function at  $\omega \geq 0$ for the low-energy branch: $2/\pi\int_0^{1.5J}\mathrm{d}\omega A^p(\mathbf{k},\omega)$
    and high-energy branch: $2/\pi\int_{1.5J}^{4J}\mathrm{d}\omega A^p(\mathbf{k},\omega)$ are shown in (a) and (b), respectively;   
    The total momentum distribution of the holes, $n_{\mathbf{k}}^{h}(2h)$, for the two-hole ground state as the sum of (a) and (b), and its derivative in momentum space, are shown in (c) and (d), respectively. 
    }
    \label{fig:nkh2h}
\end{figure}

%%%%%%%%%%%%%%%%%%%%%%%%%%%%%%%%%%%%%%%%%%%%%%%%%%%%%%%%%

\subsection{The single-particle spectral function in momentum-energy space}

The real-space information finds its correspondence in the momentum space spectrum as well, as displayed in Fig.~\ref{fig:arpes}. 
Calculations of the momentum-space spectral functions, $A^p({\bf k}, \omega)$ and $A^n({\bf k}, \omega)$, are essentially the same as the real space ones,  $A_{ii}^p(\omega)$ and $A_{ii}^n(\omega)$, with the $c_{i\sigma}$ operator in Eqs.~\eqref{eqn:Ap} and \eqref{eqn:Am} replaced by $c_{{\bf k}\sigma}$
\begin{equation} \label{eqn:Akp}
    A^p(\mathbf{k},\omega) = -\mathrm{Im} \sum_n\frac{\left|_{1h}\langle\Psi^c(n)|c^\dagger_{\mathbf{k}\uparrow}|\Psi_G\rangle_{2h}\right|^2}
    {\omega-[E_{1h}(n)-E^G_{2h}-\mu_p] +i\eta}~,
\end{equation}
\begin{equation} \label{eqn:Akm}
    A^n(\mathbf{k},\omega) = -\mathrm{Im} \sum_n\frac{\left|_{1h}\langle\Psi^c(n)|c_{\mathbf{k}\downarrow}|\phi_0\rangle\right|^2}
    {\omega +[E_{1h}(n)-E^G_{0}-\mu_m]+i\eta}~.
\end{equation}
The calculated results for both the positive/negative biases are presented in Fig.~\ref{fig:arpes} (a)-(d), respectively.

On the positive bias side [Figs.~\ref{fig:arpes} (a) and (b)], one sees that the low-energy branch of the double-peak structure in the STS spectral function (cf. Fig.~\ref{fig:stm_ii}) actually corresponds to a nodal quasiparticle like excitation with a vanishing gap and spectral weight along the nodal line and a maximum gap near the antinodal region. Especially the whole dispersion is approximately symmetric to the single-hole spectrum on the negative bias side [Figs.~\ref{fig:arpes} (c) and (d)]. 

Besides, on the positive bias, corresponding to the upper branch in the STS spectral function (cf. Fig.~\ref{fig:stm_ii}), $A^p({\bf k}, \omega)$ shows a high-energy dispersion with minima at $(\pm\pi/2,\pm\pi/2)$, 
which has been also seen in other numeric methods in the low-doping limit, e.g. the cluster theory method result of a Hubbard model \cite{Kohno2012,Sakai2018}.  
%Given both the single-hole and two-hole wave functions, a simple intuitive understanding is ready.

%As to be detailed below, the low-energy branch is characterized by a $d$-wave quasiparticle excitation while the high-energy branch comes from the ``twisted'' quasiparticle excitation with a gap $\Delta_{pg}\simeq E_{\mathrm{pair}}$ by breaking up the local tightly-bound pair of the two-hole state ~\cite{Zhao2022}.
%This ``spectrum shift'' therefore indicates that the nodal $c_{i\sigma}$ quasi-particle within low-energy branch,  compared to the nodal $\tilde{c}_{i\sigma}$ quasi-particle within the high-energy branch,  
%has smaller overlap with the two-hole ground state, 
%which is 

%\JY{\sout{except for the disappearance of the portion along the diagonal scan between $(0,0)$ and $(\pi,\pi)$ due to the $d$-wave symmetry %\cite{Zhao2022}} in the two-hole ground state. 
%On the other hand, the higher energy branch mimics the behavior of a ``twisted'' single-hole excitation with the band bottom at $(\pi/2,\pi/2)$. 

%\section{
%{\sout{The internal structure of the single-hole excitation.}
%Two-component structure in single-hole excitations }

%\subsection{Momentum distributions} % of the two-component structure}

%To get more implications of the physics at the two distinct branches, we integrate the positive bias spectrum in momentum space
To further examine the momentum structure of the two-hole ground state, one may integrate $\omega$ at $\omega\geq 0$ in the spectral function $A^p({\bf k}, \omega)$ around two branch energy scales between $0\sim 1.5J$ and $1.5J\sim 4J$, respectively, 
as shown in Fig.~\ref{fig:nkh2h} (a) and (b). 
For the low-energy branch in Fig.~\ref{fig:nkh2h} (a), 
%which corresponds to a nodal quasiparticle $c_{i\sigma}$,it has a high-density distribution near the nodal region but vanishes along the nodal direction. 
the momentum distribution vanishes along the nodal directions, which looks
%as a $d$-wave pairing order parameter, and reaches its maximum at some point between the nodal and antinodal directions.  
very similar to that of the $d$-wave pairing order parameter of the two-hole ground state in Eq.~\eqref{eqn:twoholeansatz} (cf. Fig.~3 of Ref.~\cite{Zhao2022}).

On the other hand, for the high-energy branch, the momentum distribution shows pockets around the nodal regions at $(\pm\pi/2,\pm\pi/2)$ [also cf. Figs.~\ref{fig:arpes} (a) and (b)]. 
%Combined with , one can conclude that the high-energy branch a
These pockets, with an energy scale around $\Delta_{pg}\sim 2J$, consistent well with the excitation energy $E_{pair}\simeq 1.97 J$ of the ``twisted'' quasiparticles created by breaking the bound-energy in the two-hole ground state ~\cite{Zhao2022}.

Combining the low-energy and high-energy branches, one can recover the single-hole momentum distribution $n_{\mathbf{k}}^h(2h)$ for the two-hole ground state in Eq.~\eqref{eqn:twoholeansatz}, given by
%According to a sum rule 
\begin{equation}\label{eqn:sumrule}
    2\int_0^\infty \mathrm{d}\omega A^p(\mathbf{k},\omega)/\pi \equiv  n_{\mathbf{k}}^h(2h) + \mathrm{const.}
\end{equation}
The calculated $n_{\mathbf{k}}^h(2h)$ is shown in Fig.~\ref{fig:nkh2h}(c). 
%Considering an approximate size of $4a_0\times 4a_0$ for the two-hole pair, one may roughly treat $n_{\mathbf{k}}^h(2h)$ in Fig.~\ref{fig:nkh2h}(c) as if it is a momentum distribution at finite doping $\delta\sim 1/8$. 
Note that it is a commonly used method in finite-size numerical study to utilize the gradient of the $n_{\mathbf{k}}^h(2h)$ to estimate the equiv-energy-contour shape. As shown in Fig.~\ref{fig:nkh2h}(d), the corresponding pattern of $|\nabla n_{\mathbf{k}}^h(2h)|^2$ resembles the contribution of nodal Bogoliubov quasiparticles. 
%As $n_k^h$ is expected to be smooth before and after the superconducting transition, 
%it is appealing to guess that the high energy branch will be pulled down to zero as doping concentration increases and AFLRO gets killed. 
%More discussion on the finite doping physics is given in Sec.??

%Note that due to the energy cutoff in the integration and incomplete of the wave function ansatz Eq.~\eqref{eqn:singleholeansatz_v}, 
%as well as that $n_k^h(2h)$ is estimated by fixing $i_0$ in $c^\dagger_{i_0}c_{j}$ in an OBC lattice, 
%the sum rule Eq.~\eqref{eqn:sumrule} is not exactly recovered here, especially for the high energy part around the nodal direction.  
%However, due to the high resemblance pattern between the two distinct methods, we believe the key physics are already captured. 

%The low-energy .. anti-nodal. Higher energy .. pocket. 
%While their summation, around the nodal line. 

\begin{figure*}[t]
    \centering
    \includegraphics[width=1\textwidth]{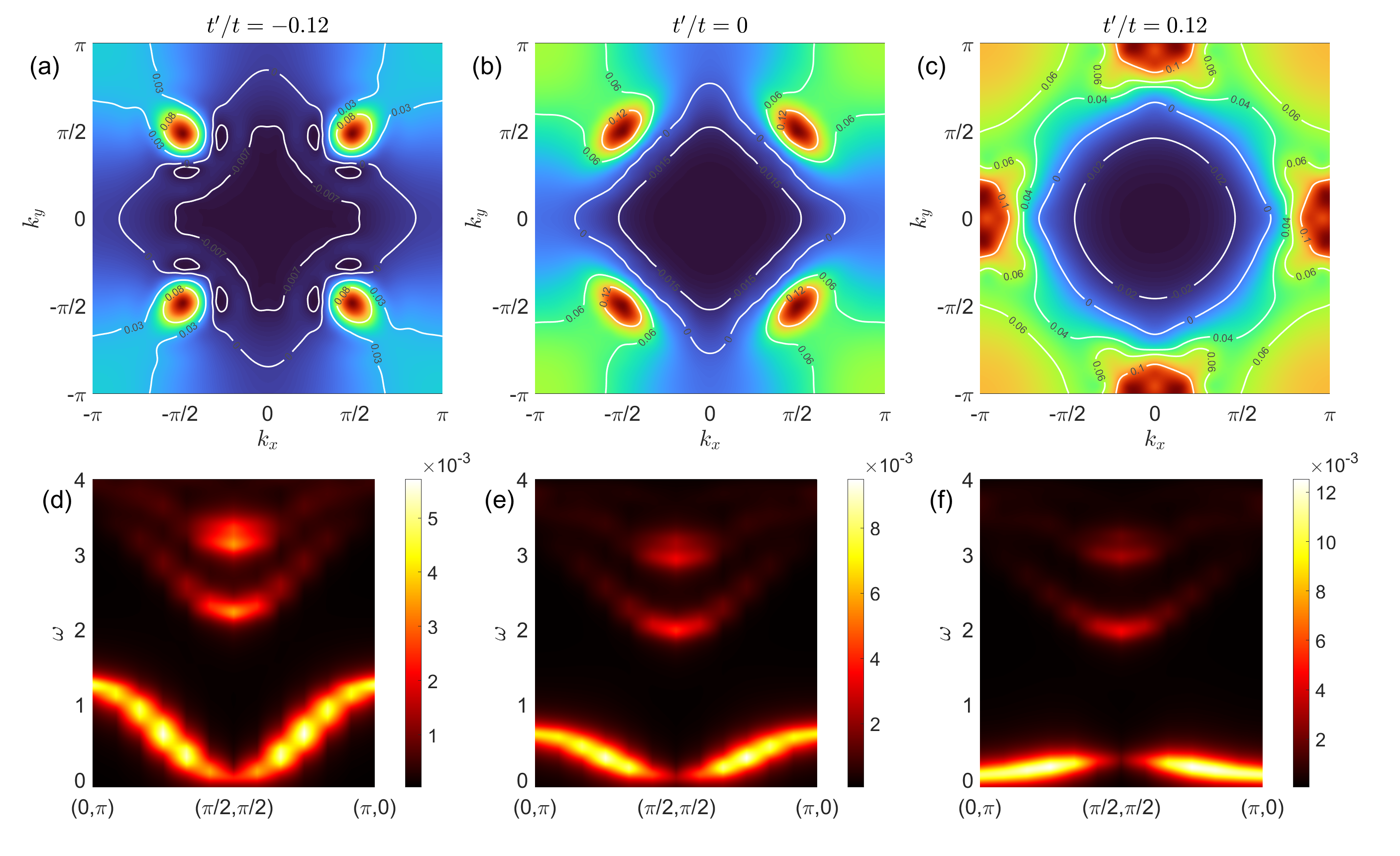}
    \caption{The hole momentum distribution $n_{\mathbf{k}}^h(1h)$ determined based on the single-hole ground state in Eq.~\eqref{eqn:singleholeansatz_v} (the half-filling background always contributes to a constant $1$ to the electron momentum distribution $n_k^e$)
    %calculated based on the integrated spectral function at $\omega<0$ % (JY: integral spectral of $\omega<0$ gives the half-filling density, which is uniform) }
        at (a) $t'/t = -0.12$, (b) $t'/t=0$ and (c) $t'/t = 0.12$;
        The corresponding spectral function at $\omega>0$ along the momenta $(\pi,0)$ to $(0,\pi)$ direction is shown for 
        (d) $t'/t = -0.12$, (e) $t'/t=0$ and (f) $t'/t = 0.12$. }
    \label{fig:nk1htp}
\end{figure*}

\subsection{Distinct effects of the NNN hopping $t'$}

It is believed that in a realistic cuprate compound, a longer-range hopping such as the NNN hopping $t'$ needs
also to be taken into account in order to describe correctly the behavior of the single-hole excitation \cite{Wells1995, Nazarenko1995, Tohyama2000}. In the following, we examine the effect on the spectral function by adding the $t'$-term given by   
\begin{equation}
    H_{t'} = -t'\sum_{\langle\langle ij\rangle\rangle,\sigma}(c^\dagger_{i\sigma}c^{}_{j\sigma} + h.c.)~,
\end{equation}
where $\langle\langle ij\rangle\rangle$ denotes $i,j$ belonging to the NNN sites. 

Previously, the effect of the NNN hopping on the two-hole pairing wavefunction in Eq.~\eqref{eqn:twoholeansatz} has been already studied in Ref.~\onlinecite{Zhao2022}. 
It is found that so long as $|t'/t|\ll 1$, the two-hole wavefunction is not obviously affected, 
except that the pairing strength monotonically grows from $t'/t<0$ to $t'/t>0$, consistent with the ED result \cite{Zhao2022}. 
%A simple understanding based on the phase string sign structure is also given there. 
%exp \cite{Wells1995,Sheng1996}
%\cite{Vishik2010,Ronning1998,Hu2018,Gao2020}. 
%SCBA \cite{Nazarenko1995, Belinicher1996}
%ED \cite{Eder1997,Kim1998}
%systematic \cite{Tohyama2000}
So the effect of the NNN term $t'$ on the single-electron spectral function will be mainly determined by the single-particle spectrum for the single-hole wavefunction in Eq.~\eqref{eqn:singleholeansatz_v}. 
%by making use both of the single-hole and two-hole wave functions Eqs.~\eqref{eqn:singleholeansatz_v} and \eqref{eqn:twoholeansatz}, 
%provided that $|t'|$ is significantly smaller than $t$ so that the phase-string effect remains dominant.

In Figs.~\ref{fig:nk1htp}(a)-(c), the hole momentum distribution $n_{\mathbf{k}}^h(1h)\equiv 1-n_{\mathbf{k}}^e(1h)$ [with $n_{\mathbf{k}}^e(1h)=\sum_{\sigma}\langle c^{\dagger}_{\mathbf{k}\sigma}c_{\mathbf{k}\sigma}\rangle $ where the average $\langle ... \rangle $ is over the single-hole state of Eq.~\eqref{eqn:singleholeansatz_v}], is calculated based on Eq.~\eqref{eqn:singleholeansatz_v} at different $t'/t$'s. The corresponding single-particle spectral function on the positive bias side is plotted along the line connecting the antinodal points in Figs.~\ref{fig:nk1htp}(d)-(f). 

At $t'/t=0$, $n_{\mathbf{k}}^h(1h)$ shows a quasiparticle peak located around at momentum $(\pm \pi/2,\pm \pi/2)$ in Fig.~\ref{fig:nk1htp} (b). 
The contour of $n_{\mathbf{k}}^h(1h)$ shows an anisotropic shape, with a much higher gradient along the nodal direction than along the antinodal direction [cf. the dispersion in Fig.~\ref{fig:arpes}]. By contrast, at $t'/t=-0.12$, the contour plot of $n_{\mathbf{k}}^h(1h)$ becomes much more isotropic, consistent with the increasing velocity along antinodal direction in Fig.~\ref{fig:nk1htp} (d). Note that a strong anisotropic dispersion around the Fermi points $(\pm \pi/2,\pm \pi/2)$ seems inconsistent with the ARPES experiments \cite{Sheng1996,Ronning1998,Hu2018,Gao2020}, where an isotropic dispersion has been observed for the cuprate materials near the half-filling. Such a discrepancy has been generally considered as a result of a negative $t'/t$ effect 
\cite{Wells1995,Vishik2010,Ronning1998,Hu2018,Gao2020,Nazarenko1995, Belinicher1996,Eder1997,Kim1998,Tohyama2000}. 
%where $t'/t\sim-0.35$ is generally used to recover the experimental dispersion. %The wave function ansatze Eqs.~\eqref{eqn:singleholeansatz}, \eqref{eqn:singleholeansatz_v} and \eqref{eqn:twoholeansatz} 
%As shown in Fig.~\ref{fig:nk1htp} (a), 
%quasi-particle dispersion around the nodal line becomes isotropic, which aligns with results from ARPES studies. 

%{\color{red}It seems our positive bias side spectrum can be compared with the electron doped APRES, such as \cite{Hu2021}.}

At $t'/t>0$, the contours of $n_{\mathbf{k}}^h(1h)$ around the Fermi points $(\pm \pi/2,\pm \pi/2)$ first become even more anisotropic, 
and then the Fermi points shift from nodal direction to the antinodal regions at $(\pm \pi,0)$ and $(0,\pm\pi)$ as shown in Fig.~\ref{fig:nk1htp} (c). Such antinodel quasiparticle excitations match well with the experimental observation in the electron-doped cuprates \cite{Hu2021}. 
%Experimentally, $t'/t>0$ corresponds to electorn doped materials up to a partile-hole transformation. 
%Note that different from the hole doped case corresponding to $t'/t\leq 0$, 
%the positive bias spectrum of $t'/t>0$, after a particle-hole transformation and a $(\pi,\pi)$ momentum shift, 
%can be directly compared with ARPES experiments on electron doped materials. 
%The two-component structure also shows interesting consistency with the experimental observed two-component structure in the electron doped materials\cite{Hu2021}. 

%Our calculation indicates the physics of these two pockets should be distinct. 

Finally, we make an important observation at $t'/t>0$. Despite the significant changes in low-energy branch single-particle dispersion as a function of $t'/t$, the dispersion of the high-energy branch remains almost unaffected as $t'/t$ varies, as shown in Figs.~\ref{fig:nk1htp} (d)-(f). As discussed in the previous section, the high-energy branch of the local spectral function is dominant with the physics of a ``twisted'' hole described by $\tilde{c}_{i\sigma}$. The insensitivity of such a high-energy dispersion indicates that the ``twisted'' quasiparticle physics is robust under small perturbation of $|t'/t|$. Namely the ``twisted'' particles in the $t$-$J$ model or doped Mott insulators  plays the most essential building-block role, independent of doping concentration, AF correlation length, and small band structure perturbations such as $t'$, at least in the small doping regime. Experimentally a two-band structure has been indeed observed \cite{Hu2021} by ARPES on the electron doped side, with the low-lying branch (Fermi points) lying along the antinodal regions, 
while the high-energy branch remains along the nodal direction with a gap $\sim 0.3 \mathrm{eV}$ . 
%consistent with our physical intuition that the is determined by a singular fluctuating $\pm$ sign structure. 

%\subsection{\color{teal} Zhang-Rice Singlet?}

%\subsection{Indications for finite doping physics}

%\JY{
%Here, the ``twisted'' particle part is already well studied in the former papers in both the single-hole and two-hole doped case. 
%On the other hand, the behavior of the anti-phase factor has only been discussed in the finite doping theory, 
%where it is assumed to be smeared/condensed to become uniform flux $\delta\pi$. 
%This flux will self-consistently turns the spinon into a short ranged ordered state. 
%it can be either condensed or confined depending on the spin-spin correlations on the AF background. 
%At finite doping where the AFLRO is killed by the condensation of holons, the phase factor is condensed. 
%the superconductivity phase coherence is constructed. 
%Ref.~\ref{Jianghao&jiaxin} also shows that the phase factor is important for the calculation of the Fermi arc.

%3. localized/smeared. 

\section{Summary and Discussion}\label{sec:conclusion}

The STS and ARPES experiments are measurements based on the single-electron probe. The physical interpretation of the experimental results, i.e., the single-electron spectral function, has been well established for weakly-interacting electron systems in the framework of the Landau quasiparticle description. However, how to understand the data in a strongly correlated system like the cuprate remains a challenge. In the present paper, the single-electron spectral function for a two-hole-doped ground state in the $t$-$J$ model has been systematically explored, including the effect of the NNN hopping $t'$. The results are quite nontrivial, which reveal some important features solely associated with the strongly correlated nature of the ground state. 

The spectral function at $\omega>0$, $A^p(\omega)$,  
measures the overlap of the two-hole ground state %[Eq.~(\ref{eqn:twoholeansatz})] 
with a single-hole excitation %$|\Psi^c(n)\rangle_{1h}$ [Eq.~(\ref{eqn:singleholeansatz_v})] 
via injecting a bare electron in STS, i.e.,
\begin{equation}\label{overlap}
\propto\left|_{1h}\langle\Psi^c(n)|c^\dagger_{i\uparrow}|\Psi_G\rangle_{2h}\right|^2~,
\end{equation}
where the excitation energy of $|\Psi^c(n)\rangle_{1h}$ is determined by VMC, which then gives rise to a double-peak structure as shown in Fig.~\ref{fig:stm_ii}. Generalized to momentum space, the double-peak structure further extends into two branches of distinctive dispersions in $A^{p}({\bf k}, \omega)$ as shown in Figs.~\ref{fig:arpes} (a) and (b). The high-energy mode is shown to have a threshold at $\Delta_{pg}\simeq E_{pair}$, i.e., the pair gap associated with the $s$-wave pairing of the twisted holes in the ground state \cite{Zhao2022}. The main weight of such a mode concentrates around momentum points, $(\pm\pi/2,\pm\pi/2)$, and extends along the diagonal direction towards $(\pm\pi,\pm\pi)$ as shown by the momentum distribution in Fig.~\ref{fig:nkh2h}(b). In particular, such a high-energy branch remains barely changed by the NNN $t'$, %with the minimal points at $(\pm\pi/2,\pm\pi/2)$, 
in contrast to the sensitivity of the low-energy branch under the influence of $t'$ as shown in Figs.~\ref{fig:nk1htp} (d)-(f).

By contrast, the low-lying branch shown in Fig.~\ref{fig:stm_ii} corresponds to a dispersion extending from  the nodes $(\pm\pi/2,\pm\pi/2)$ towards the antinodal regions as illustrated in  Figs.~\ref{fig:arpes} (b), while the spectral weight  vanishing at the nodes  and along the nodal directions in  Figs.~\ref{fig:arpes} (a) and (b), which is consistent with the $d$-wave-like pairing symmetry of the ground state wavefunction \cite{Zhao2022}. Interestingly, a symmetric low-energy peak is also found at $\omega<0$ in $A^n(\omega)$  (cf. Fig.~\ref{fig:stm_ii}). In momentum space, such a low-lying dispersion at $\omega<0$ also matches well with that at $\omega>0$ along, e.g., the $(0,\pi)$ and $(\pi,0)$  (cf. Fig.~\ref{fig:arpes}), except for the diminishing weight along the diagonal nodal lines in the latter as mentioned above. In particular, the energy dispersion at $\omega<0$ agree remarkably well with the previous QMC result (white cross) in the whole Brillouin zone as shown in Figs.~\ref{fig:arpes} (c) and (d). Furthermore, the anisotropic ``Dirac-like'' dispersion around $(\pm\pi/2,\pm\pi/2)$ can be further tuned towards more isotropic at $t'/t<0$, whereas the Fermi points can be even shifted to the antinodal regions at $(\pm\pi,0)$ and $(0,\pm\pi)$ at a finite $t'/t>0$ (cf. Fig.~\ref{fig:nk1htp}).  

What can one read off from these results? Previously it has been shown \cite{Zhao2022} that the two-hole ground state $|\Psi_G\rangle_{2h}$ has a dichotomy in its hole pairing structure: a $d$-wave Cooper pairing embedded in a stronger $s$-wave pairing of ``twisted'' holes. The present single-electron spectral functions faithfully reveal such a composite pairing structure, with two distinct energy scales representing the pairing at different length scales.  In particular, the pairing strength is essentially derived from the latter. This is consistent with the present result that the low-lying $d$-wave-like spectrum in $A^p(\omega)$ is essentially symmetric with regard to that shown in $A^n(\omega)$, whereas the latter only probes a single-hole state without pairing.   On the other hand, while the double-peak structure at $\omega>0$ qualitatively explains the STS meansurement \cite{Li2022}, the high-energy branch of the single-hole excitation is clearly absent at $\omega<0$ in $A^n(\omega)$. This is  also consistent with the STS \cite{Li2022}, but the underlying mechanism is quite unconventional as to be discussed below.

\begin{figure}
    \centering
    \includegraphics[width=0.45\textwidth]{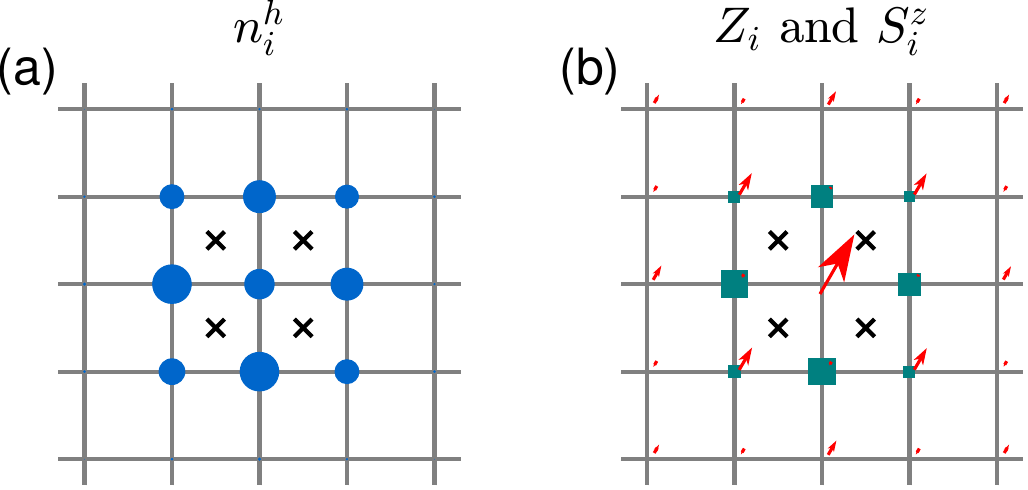}
    \caption{``Orthogonality catastrophe'' effect in a localized single-hole state, which may give rise to the four-lobe clover pattern seen in STS \cite{Ye2023b}. (a) The hole density profile (blue dots); (b) The corresponding quasiparticle spectral weight $Z_i$ (teal squares) and the spin distribution $S^z_i$ (red arrows). Here the wavefunction is constructed according to Eq.~(\ref{eqn:singleholeansatz_v}) with the center (cross) of the anti-vortex restricted within a $2\times 2$ lattice region to simulate a self-trapping by impurity.  }
    \label{fig:zrs}
\end{figure}

\subsection{``Orthogonality catastrophe''}

%Finally, we would like to point out that the vanishing \emph{high-energy } branch in $A^n(\omega)$ at $\omega<0$ and the vanishing spectral weight in the \emph{low-energy} branch along the nodal lines in $A^p(\omega)$ at $\omega>0$ are all enforced via the destructive quantum interference of the phase string factor in Eq.~(\ref{eqn:phasestringoperator}). It leads to the so-called  ``orthogonality catastrophe'' effect as to be further discussed below.       

Finally, we discuss an important issue about how to correctly interpret an experimental observable in a strongly correlated system. In the present work, the spectral function measures the overlap of the single-hole excitation $|\Psi^c\rangle_{1h}$ with either the two-hole ground state or the half-filling ground state, with injecting an electron at $\omega>0$ and a bare hole at $\omega<0$, respectively. At $\omega<0$, we have seen that the high-energy single-hole excitation has no contribution to $A^n(\omega)$. On the other hand, for the low-energy branch, $A^p(\omega)$ at $\omega>0$ vanishes along the nodal lines in contrast to the full single-hole dispersion in $A^n(\omega)$ at $\omega<0$. All of these can be attributed to the meron-vortex or phase-string field appearing in Eq.~(\ref{eqn:singleholeansatz_v}), which leads to the diminishing overlap.

As a matter of fact, the pairing symmetry of the two-hole state was previously examined \cite{Zhao2022} by a finite overlap with a Cooper pair state of the conventional quasiparticles, $c_{{\bf k}\uparrow}c_{{\bf -k}\downarrow}|\phi_0\rangle$, where a $d$-wave symmetry is realized also by the destructive interference of the vortex field even though  the amplitude $g(i,j)$ has an $s$-wave symmetry in Eq.~(\ref{eqn:twoholeansatz}) . %Thus, the $s$-wave pairing of the ``twisted'' holes in Eq.~(\ref{eqn:twoholeansatz}) 
Namely even though the ground state is actually an $s$-wave pairing of the ``twisted'' holes, which are surrounded by transverse spin currents when they are not paired, it looks like the $d$-wave pairing by the quasiparticle probe, resulting in the so-called $d$- and $s$-wave dichotomy. Here the $s$-wave pairing of the ground state is like a ``dark matter''. Without the latter, the pairing of the quasiparticles alone would be rather weak as shown in the VMC calculation \cite{Zhao2022}. How to reveal such ``dark matter'' in an experimental measurement, however, is highly nontrivial due to the ``orthogonality catastrophe'' or the destructive interference effect by the phase-string field as discussed above. In the present study, its appearance as the precursor of a pseudogap behavior is shown in $A^p(\omega)$.   

Now we show another example that a strongly correlated ``dark matter'' component may not be directly revealed in an STS spectral function. A doped hole may be trapped by an impurity in the lightly-doped cuprate. Instead of directly simulating the local impurity effect \cite{Ye2023b,Xia2024}, one may construct a localized single-hole state $|\Psi_{im}^{c}\rangle_{1h}$ based on Eq.~(\ref{eqn:singleholeansatz_v}) by fixing the location of the antivortex within the central four plaquettes as indicated by the crosses in Fig.~\ref{fig:zrs}. The distribution of the localized hole is illustrated Fig.~\ref{fig:zrs} (a). The corresponding spectral weight of such an impurity state defined by
\begin{equation} \label{eqn:Apfourlobe}
    Z_{i} \equiv \left|\langle\phi_0|c^\dagger_{i\downarrow}|\Psi_{im}^{c}\rangle_{1h}\right|^2~,
\end{equation}
is shown in Fig.~\ref{fig:zrs} (b), together with an average spin distribution (extra spin of $S^z=1/2$ is introduced by creating such a hole). The sharp contrast between Fig.~\ref{fig:zrs} (a) and Fig.~\ref{fig:zrs} (b) clearly illustrates how a partial information about the (approximately uniform) local distribution of the hole would be absent in the local STS probe \cite{Ye2023b} to give rise a four-lobe clover pattern in the latter, again thanks to the destructive interference effect by the phase-string vortex phase in $|\Psi_{im}^{c}\rangle_{1h}$.   

Similarly, as previously discussed in the DMRG and VMC studies \cite{Zhu2018a,Zhao2023} of the single-hole-doped two-leg ladder system, the $Z_i\neq 0$ ``quasiparticle'' component only constitutes an ``insignificant'' part of the ground state, in which the dominant component is composed of the doped hole accompanied by a cloud of transverse spin currents created by the phase-string vortex [cf. Eq.~(\ref{twisted})]. The latter has no contribution to $Z_i$ and may be regarded as a ``dark matter'' beyond the single-electron probe by STS and ARPES. This is because the spin currents created in the half-filled spin background cannot be registered directly in the single-electron spectral function. One may easily understand this by noting that $n^e_{\bf k}\equiv 1$ at half-filling no matter what the spin excitations are. Of course, an indirect evidence of the incoherence of the hole may be still shown in the spectral function by a broad momentum distribution of the doped hole due to a partial momentum carried away by the surrounding spin currents in the background \cite{Zhu2018a,Zhao2023}.

At a finite but small doping concentration, we have pointed out that the ground state may be reasonably regarded as the Bose condensation of the tightly-bound hole pairs as described by Eq.~(\ref{eqn:SC}). Here the present scheme is expected to still work approximately, except that the overall superexchange coupling $J$ in $\Delta_{pg}$ should be reduced to an effective $J_{\mathrm{eff}}$ which monotonically decreases with doping \cite{Ma2014}. In this case, 
%both $|\Psi^c\rangle_{1h}$ and $|\Psi_{1h}\rangle$ can reach a finite overlap 
due to the phase coherence reached in Eq.~(\ref{Omega}), the low-branch excitations in Fig.~\ref{fig:stm_ii} at both $\omega>0$ and $\omega<0$ can truly become the coherent nodal Bogoliubov quasiparticle excitations. More elaborate understanding based on Eq.~(\ref{eqn:SC}) at finite doping is presented elsewhere \cite{Zhang2020,Zhang2022}.

\begin{acknowledgments}
%\textit{Acknowledgements.---}
We acknowledge stimulating discussions with Yayu Wang, Hai-Hu Wen, and Jia-Xin Zhang. This work is supported by MOST of China (Grant No. 2021YFA1402101) and NSF of China (Grant No. 12347107).
\end{acknowledgments}

%\bibliography{ref/nameAbrv.bib,ref/refs.bib}
\bibliography{ref/nameFull.bib,ref/refs.bib}

\end{document}